\documentclass[12pt,preprint]{aastex}
\usepackage{epsf}

\newcommand{\be}{\begin{displaymath}}
\newcommand{\ee}{\end{displaymath}}
\newcommand{\bea}{\begin{eqnarray}}
\newcommand{\eea}{\end{eqnarray}}

\shortauthors{Denissenkov \& Pinsonneault}
\shorttitle{The impact of carbon enhancement on extra mixing}

\begin{document}

\title{THE IMPACT OF CARBON ENHANCEMENT ON EXTRA MIXING IN METAL-POOR STARS}

\author{Pavel A. Denissenkov\altaffilmark{1,2}, and Marc Pinsonneault\altaffilmark{1}}
\altaffiltext{1}{Department of Astronomy, The Ohio State University, 4055 McPherson Laboratory,
       140 West 18th Avenue, Columbus, OH 43210; dpa@astronomy.ohio-state.edu, pinsono@astronomy.ohio-state.edu.}
\altaffiltext{2}{On leave from Sobolev Astronomical Institute of St. Petersburg State University,
   Universitetsky Pr. 28, Petrodvorets, 198504 St. Petersburg, Russia.}
 
\begin{abstract}
We critically examine the constraints imposed by carbon-enhanced metal-poor (CEMP) stars on the
mixing mechanisms that operate in red giants.  CEMP stars are created when the surface layers of a metal-poor dwarf are
enriched with He-burning products via mass transfer from an evolved donor.  The difference between
main sequence (MS) and red giant CEMP abundances can be used as a diagnostic of the timescale
for the mixing of the processed material into stellar interiors on the MS.  Abundance trends
with luminosity among red giant CEMP stars test theories of canonical extra mixing for low mass giants with
a high bulk metallicity.  We find a significant dilution in CN enrichment in giant CEMP stars relative to their MS
precursors, and take this as evidence that thermohaline mixing induced by mean molecular weight inversions is
ineffective in CEMP stars.  This contradicts models that rely on efficient thermohaline mixing induced by small $\mu$
gradients in red giants, because such models would predict that MS CEMP stars with large $\mu$ inversions
would be homogenized on a very short timescale.  The data do not rule out slower MS
thermohaline mixing comparable to previously published estimates.  We also find that canonical extra mixing is strongly
suppressed in CEMP giants relative to stars with the same iron abundance.  A likely cause is the shift in the location
of non-equilibrium CN processing relative to the steep $\mu$ gradient in the hydrogen burning shell, which also occurs in
solar metallicity RGB stars.  Implications for the mass accreted by CEMP stars and the mechanism for canonical extra mixing are discussed.
\end{abstract} 

 
\section{Introduction}
\label{sec:intro}

It has long been known that asymptotic giant branch (AGB) stars can have their surfaces enriched in helium, carbon and slow neutron capture products, and such stars are known
nucleosynthetic sources (e.g., \citealt{h05}).  If there is a nearby main sequence (MS) companion, mass transfer from the giant can produce surface abundance anomalies that persist even after
the AGB star is gone.  Such a system would appear to be a chemically peculiar MS star.  In recent years there have been large-scale surveys
for metal-poor stars that have included substantial spectroscopic follow-up campaigns, such as the Hamburg/ESO R-process Enhanced Star Survey (HERES: Christlieb et
al. 2004).
A large population of carbon-enhanced metal-poor (CEMP) stars has been discovered (e.g., \citealt{bch05,lea06}),
and a majority of the CEMP stars have white dwarf companions (\citealt{lea05}).  This is strong evidence that there is a significant population of
low mass metal-poor stars which have experienced mass transfer from a former AGB companion.  Much of
the interest in these stars has been driven by chemical evolution studies (e.g., \citealt{fea05,t07}).
However, we wish to emphasize another important property of CEMP stars: they
are unique astrophysical laboratories that can be used to test theories of mixing in low mass stars.  A comparison of MS and evolved CEMP stars can test
the timescale for mixing on the MS, and abundance trends within evolved stars can be used
to test mixing on the red giant branch (RGB).  In this paper we demonstrate that
CEMP stars are not fully homogenized on the MS and that they do not experience the same RGB mixing pattern as other stars with the same iron abundance, and
explore the theoretical explanations and consequences of both results.

There are sound theoretical reasons to expect internal mixing in stars that accrete mass from an evolved companion.  Metal-poor turnoff stars have thin
surface convection zones, and as a result their surface abundances after mass transfer will
reflect the composition of the accreted material.  The He and C rich ejecta of an evolved AGB star has a higher mean molecular weight $\mu$ than a normal stellar envelope.  Although the strong density gradients in a star prevent a classical Rayleigh-Taylor instability when such ejecta is accreted, Stancliffe et al. (2007) pointed out that this structure is unstable against thermohaline
convection\footnote{Thermohaline convection is driven by a density difference between a fluid element and its surrounding medium,
provided that the latter has a higher mean molecular weight and the heat exchange keeps a temperature difference sufficiently small.}.
The precise timescale, however, depends critically on assumptions about the fluid geometry.  Under the assumption of a simple spherical geometry (following \citealt{kea80}), Stancliffe et al. derived a
characteristic timescale of order 1 Gyr for homogenizing a CEMP star.  In this theoretical model the surface abundances are diluted through 90\% of the mass of the star within this timescale for
typical companion masses and ages, with only the nuclear processed core escaping full mixing.  As supporting evidence the authors cite both the
absence of extremely high ([C/Fe]\footnote{We use the standard spectroscopic notation: [A/B]\,=\,$\log\,(N({\rm
A})/N({\rm B}))-\log\,(N({\rm A})/N({\rm B}))_\odot$, where $N({\rm A})$ and
$N({\rm B})$ are number densities of the nuclides A and B.}\,$ > 2.5$) enrichment in MS stars and the apparent lack of a difference between MS and RGB surface abundances.

 There has also been considerable interest in thermohaline mixing as an explanation for deep mixing in the envelopes of red giants.
There is clear evidence that ordinary low-mass giants experience {\it in situ} mixing of CN-processed
material as they become more luminous (e.g., \citealt{grea00,sm03}).
Because of its universality, \cite{dv03} proposed to call it canonical extra mixing.
Rotational mixing has long been a popular explanation (e.g., \citealt{sm79,chea98,dt00,dv03,dea06}),
but viable models have proven to be difficult to construct (\citealt{chea05,pea06}).
Attention has therefore recently shifted towards other physical mechanisms.

When performing hydrodynamic simulations using the 3D stellar evolution code {\tt Djehuty}, \cite{eea06} 
found unexpected convective-like motions in the radiative envelope of an upper RGB model.  
They attributed this to a Rayleigh-Taylor instability (RTI) driven by a mild ($\Delta\mu/\mu\approx -10^{-4}$) 
$\mu$ inversion generated by $^3$He burning.  $^3$He is produced by non-equilibrium pp burning on 
the MS and dredged up on the lower RGB.  On the upper RGB material enriched in $^3$He falls onto 
the hydrogen burning shell and the $^3$He($^3$He,\,2p)$^4$He reaction lowers $\mu$, while equilibrium 
H burning raises $\mu$ in still deeper layers; the net effect is a well-defined local minimum above the hydrogen burning shell.

We note that the RTI would not be expected in compressible, and hence stratified, stellar material.  
Instead, thermohaline convection (for the same reasons as in the CEMP case) would be a natural consequence.  
This phenomenon would be an attractive candidate to drive canonical extra mixing if it could provide 
mixing depths and rates consistent with their empirically constrained values.  
However, \cite{chz07} as well as \cite{dp07} have demonstrated that the $^3$He-driven thermohaline convection will be fast enough
to explain the RGB mixing patterns only if fluid elements have finger-like structures with a length to diameter
ratio $l/d\ga 10$.  This produces a much shorter mixing timescale than proposed by Kippenhahn et al. (1980).

If canonical extra mixing is to be explained by thermohaline convection, the inferred mixing timescale
in MS CEMP stars should be extremely short (of order 1\,--\,10 Myr).  If CEMP stars are fully mixed after
a mass transfer episode, this implies that there will be no dilution of the abundance anomalies during the classical first dredge-up (FDU).  By contrast, if mixing is
inefficient on the MS one might expect that large abundance anomalies confined to the outer layers will be diluted
when the star goes to the RGB and develops a deep
surface convection zone.  We therefore propose to use the relative abundances in MS and RGB CEMP stars as a diagnostic of the mixing timescale on the MS,
and by extension as a test of the timescale for thermohaline convection in a very favorable environment.

Canonical extra mixing might also be expected to set in for CEMP stars when they reach the upper RGB, and a distinctive manifestation
of canonical extra mixing at low metallicities is a decline of [C/Fe] with increased luminosity (e.g. Smith \& Martel 2003).
There have been published claims (\citealt{lea06,aea07})
that the abundances of CEMP stars decline with increased luminosity.  However, these studies did not separate out the MS/RGB transition (which tests the
depth of accreted material) from the RGB trends.  We therefore both examine the data for trends with luminosity and compare to theoretical expectations for empirical
models based on normal metal-poor giant data.
Again, in addition to the direct implications our test can be used to study the metallicity dependence of RGB mixing.
It is known that solar-abundance open-cluster RGB stars have weaker extra mixing than globular-cluster giants
(e.g., \citealt{chea05}). However, the open-cluster giants are both more metal-rich and more massive; both properties
could influence the mixing pattern. CEMP stars permit a direct comparison of low-mass stars with different metal abundances.

\section{Observational Data}

Ideally, our goal would be to select a sample of binary mass transfer systems where we can
look for abundance trends both between MS and RGB stars and with luminosity on the RGB.
There are now large samples of CEMP stars, but there are obvious potential biases associated
with examining abundance patterns in samples defined by abundance measurements.  As a result,
some discussion of the selection of the observational sample is essential.

One important clue is that the amount of carbon in CEMP stars is very high relative to halo non-C-enhanced stars.  This is
most heavily seen in the anti-correlation between [C/Fe] and [Fe/H] (Fig.~3 in \citealt{aea07}).  
This is the pattern that one would expect
if mass transfer from an evolved companion gave the recipient star so much C that the initial inventory was
negligible.  However, especially at the iron-poor end, some chemical evolution models might be able to produce
exotic abundance patterns for the primordial composition (e.g., \citealt{rea05,fea05,kea07}).  Fortunately, the binary mass transfer model makes other
predictions: there should still be a white dwarf companion and there should be associated s-process anomalies.
A majority of CEMP stars exhibit radial velocity variations consistent with a white dwarf companion; the observed
fraction of 68\% is consistent with all CEMP stars being in binaries (Lucatello et al. 2005). Extensive surveys
have also shown s-process enrichment (e.g., \citealt{bea05,aea07}).

We have therefore used two samples of CEMP stars: a smaller sample from \cite{aea07} in which
only stars with s-process enhancement ([Ba/Fe]\,$> +0.5$) were included, and a larger one (\citealt{lea06}) where
the standard definition ([C/Fe]\,$ > 1$) to distinguish between CEMP and normal field halo stars was used.  
Lucatello et al.'s sample is in fact a subsample of stars from the HERES survey (\citealt{bea05}) that were re-analyzed.
We did not use the stars with [C/Fe]\,$> 1$ from Barklem et al.'s original paper.
The automated procedure used in that paper to infer $T_{\rm eff}$ values from colors became unreliable
for CEMP stars with strong CH, C$_2$, and CN molecular bands (Lucatello, 2007, private communication).
The luminosities were inferred from the surface gravity and $T_{\rm eff}$ with an assumed mass of $M=0.8$\,--\,$0.85\,M_\odot$.
The values of $\log\,L/L_\odot \la 0.8$ and $\log\,L/L_\odot \ga 1.4$ were defined as the divisions between unevolved MS and subgiant stars that have not
experienced the FDU yet and evolved RGB stars in which the FDU has already ended.
We have found that 10 out of 14 unevolved stars in the second sample have [Ba/Fe]\,$> +1$ according to Barklem et al.'s data, which is a distinctive feature of
accretion of AGB material.

Under the binary mass transfer hypothesis the initial carbon inventory is dwarfed by the transferred carbon, so the physical
quantities of interest are [C/H] and [N/H] rather than [C/Fe] and [N/Fe].  As a result, we work in the former plane.  There is a striking
difference in carbon as a function of luminosity between CEMP and normal halo stars (Fig.~\ref{fig:f1}, blue open and black filled
circles, respectively), with canonical extra mixing being
clearly seen in the non-C-enhanced stars above $\log\,L/L_\odot\approx 2.2$
but not in the CEMP stars. The black curve reproduces the evolutionary decline of [C/H] in the non-C-enhanced stars by
our stellar models with canonical extra mixing (for details, see next section).
The red curve is a model constructed under the assumption that neither the MS thermohaline
convection nor RGB extra mixing have taken place. The only mixing event experienced by this model is
the standard FDU. It seems to adequately describe the [C/H] abundances in the CEMP stars.

We compare the trends in our small and large CEMP samples in Fig.~\ref{fig:f2},
and find that they are very similar.  In fact, the [C/H] abundances in the two samples have the same linear fit (\citealt{lea06,aea07}).
We also note that there is a small, but real, difference between the MS and RGB abundances.
To be more precise, \cite{lea06} have obtained very similar slopes $-0.18\pm 0.05$, $-0.21\pm 0.05$, and $-0.19\pm 0.04$
for [C/H], [N/H], and [(C+N)/H], respectively.
We argue that this is evidence for dilution of enriched material in the MS stars by the classical FDU in the RGB stars.
Both effects are quantified below.

\section{Reduced Efficiency of Canonical Extra Mixing in RGB CEMP Stars}

In this section, we show that the efficiency of canonical extra mixing in CEMP stars is strongly reduced.
This is most likely due to shallower mixing which does not penetrate deep enough to dredge up
carbon-deficient material. The most plausible interpretation of the shallower mixing depth is that it
is limited by a steeper $\mu$-gradient built up by H burning on the abundant $^{12}$C nuclei.
\cite{dea06} came to a similar conclusion for solar-metallicity
red giants. Given that RGB mixing is inefficient in CEMP stars, we can use the C and N abundance
distributions for unevolved (the MS and subgiant branch) and evolved (upper RGB) CEMP stars
to test the FDU and draw conclusions about the efficiency of thermohaline convection in MS CEMP stars.
This will be done in the next section.

The slope of [C/H] as a function of $\log\,L/L_\odot$ produced by canonical extra mixing
depends on the mixing depth and diffusion coefficient $D_{\rm mix}$.
We model canonical extra mixing by setting                 
$D_{\rm mix}$ equal to a fixed fraction of the thermal diffusivity $K$.
This parameterization takes into account that
large-scale mixing instabilities in a stably stratified radiative zone may only develop
on a timescale longer than the thermal one, in which case one would expect $D_{\rm mix} < K$. 
The known examples of such instabilities that have been proposed as possible driving mechanisms 
for canonical extra mixing are rotational shear instability (\citealt{dt00,dv03,dea06})
and thermohaline convection (\citealt{chz07,dp07}). In the latter case, 
the diffusion coefficient can be estimated using the following heuristic expression:
\bea
D_{\rm mix} \approx \left(\frac{l}{d}\right)^2\frac{|\nabla_\mu|}{\nabla_{\rm ad}-\nabla_{\rm rad}}K,
\label{eq:dmix}
\eea
where $\nabla_\mu$, $\nabla_{\rm ad}$, and $\nabla_{\rm rad}$ are the $\mu$- and temperature (adiabatic and radiative) gradients
(logarithmic and with respect to pressure). Equation (\ref{eq:dmix}) adequately represents
both the diffusion coefficient for thermohaline convection derived by \cite{kea80}
\bea
D_{\rm Kipp} = \frac{3K}{\nabla_{\rm ad}-\nabla_{\rm rad}}\,|\nabla_\mu|,
\label{eq:dkipp}
\eea
who argued that the length (or, the mean free path) $l$ of a convective element
should be of the order of its diameter $d$, and the rate of mixing by elongated narrow ``fingers'' ($l>d$)
advocated by \cite{u72}
\bea
D_{\rm Ulrich} = \frac{8}{3}\pi^2\,\frac{K}{\nabla_{\rm ad}-\nabla_{\rm rad}}\,|\nabla_\mu|\left(\frac{l}{d}\right)^2.
\label{eq:dulrich}
\eea

Unfortunately, the ratio $l/d$ cannot be estimated from first principles. Semi-empirical models
require $l/d > 10$ to generate sufficient mixing to explain the data (\citealt{chz07,dp07}). The maximum depth of the $^3$He-driven thermohaline
convection is uncertain also. The naive assumption that it coincides with
the location of the minimum on the $\mu$-profile produces mixing that is too shallow (\citealt{dp07}).
Therefore, we have chosen to specify the mixing depth $r_{\rm mix}$ in units of $R_\odot$ and adjust it appropriately.
This choice is motivated by our finding that the $\mu (r)$ does not change much with time.
In the absence of any information on how the mixing depth should vary
with time, keeping $r_{\rm mix}$ constant is a reasonable parameterization.

To allow an interpolation in the abundances of He and C as the AGB material gets redistributed
inside an accreting star by convection and extra mixing we have modified our stellar evolution code
(\citealt{dea06}) to include the 2005 version of the OPAL equation of state 
(\citealt{rea96})\footnote{The update is at {\tt http://www-phys.llnl.gov/Research/OPAL/EOS\_2005/}.}.
Taking into account that the average metallicity of the non-C-enhanced stars from the spectroscopic study of \cite{bea05}
is $\langle [\mbox{Fe}/\mbox{H}]\rangle\approx -2.7$,
we have represented them with a stellar model having the heavy-element mass fraction $Z = 4\times 10^{-5}$.
For the initial helium mass fraction, stellar mass, and mixing length parameter we have used the values $Y=0.24$, $M = 0.85\,M_\odot$, and 
$\alpha_{\rm MLT} = 1.58$, respectively.

\begin{figure}
\plotone{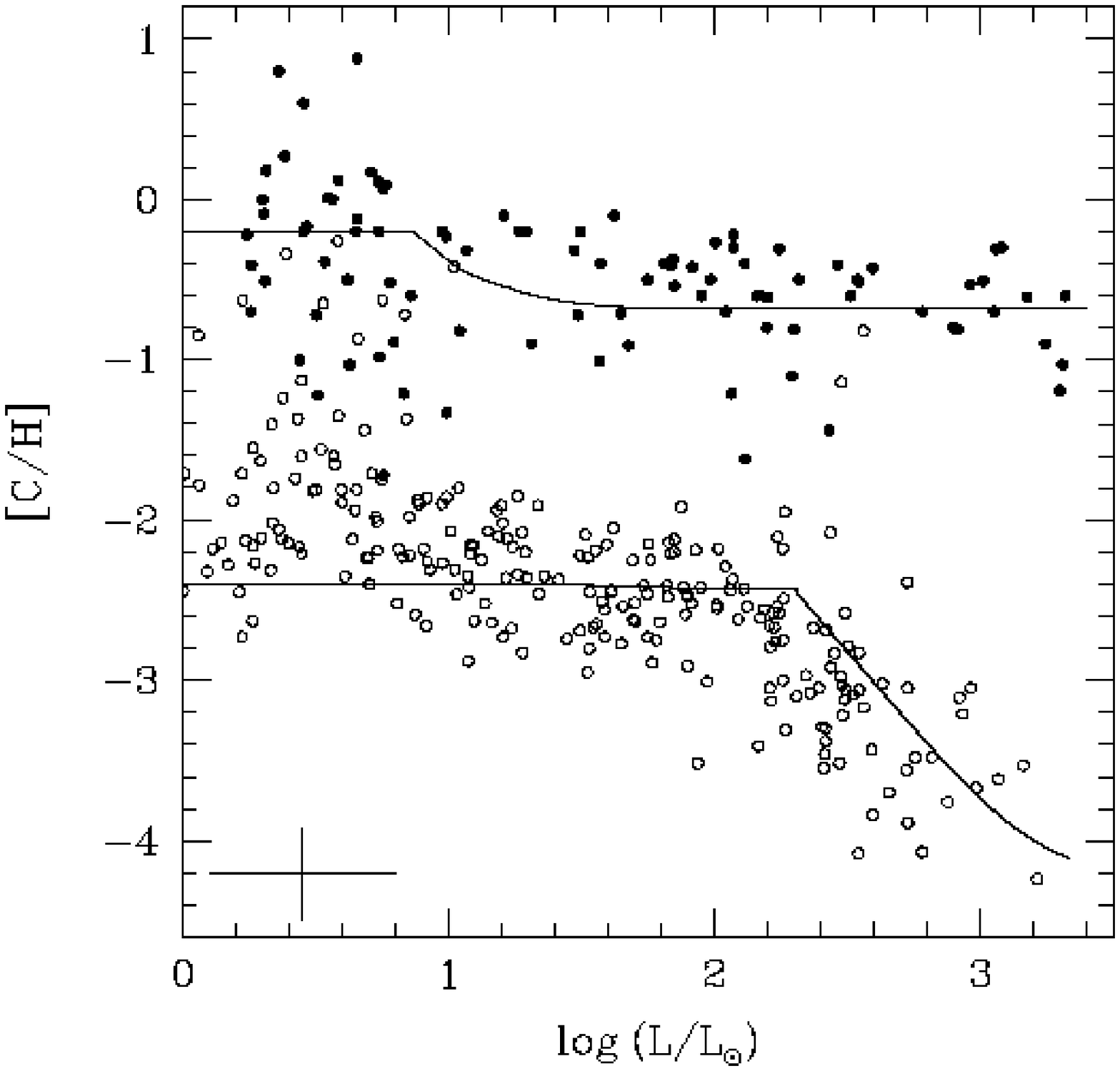}
\caption{The C abundance as a function of luminosity in the CEMP (black filled circles)
         and non-C-enhanced VMP (blue open circles; the blue cross
         shows the errorbars) stars.  The latter are seen to experience
         canonical extra mixing at $\log\,L/L_\odot \ga 2.2$. The declining part of
         the black curve has been computed for the mixing depth $r_{\rm mix} = 0.05\,R_\odot$ (panels A and C in Fig.~\ref{fig:f4})
         and diffusion coefficient $D_{\rm mix} = 0.1\,K$, where $K$ is the thermal diffusivity.
         The only mixing episode experienced by the CEMP stars appears to be the standard
         first dredge-up at $0.8\la\log L/L_\odot\la 1.4$ (red curve).
         }
\label{fig:f1}
\end{figure}

\begin{figure}
\plotone{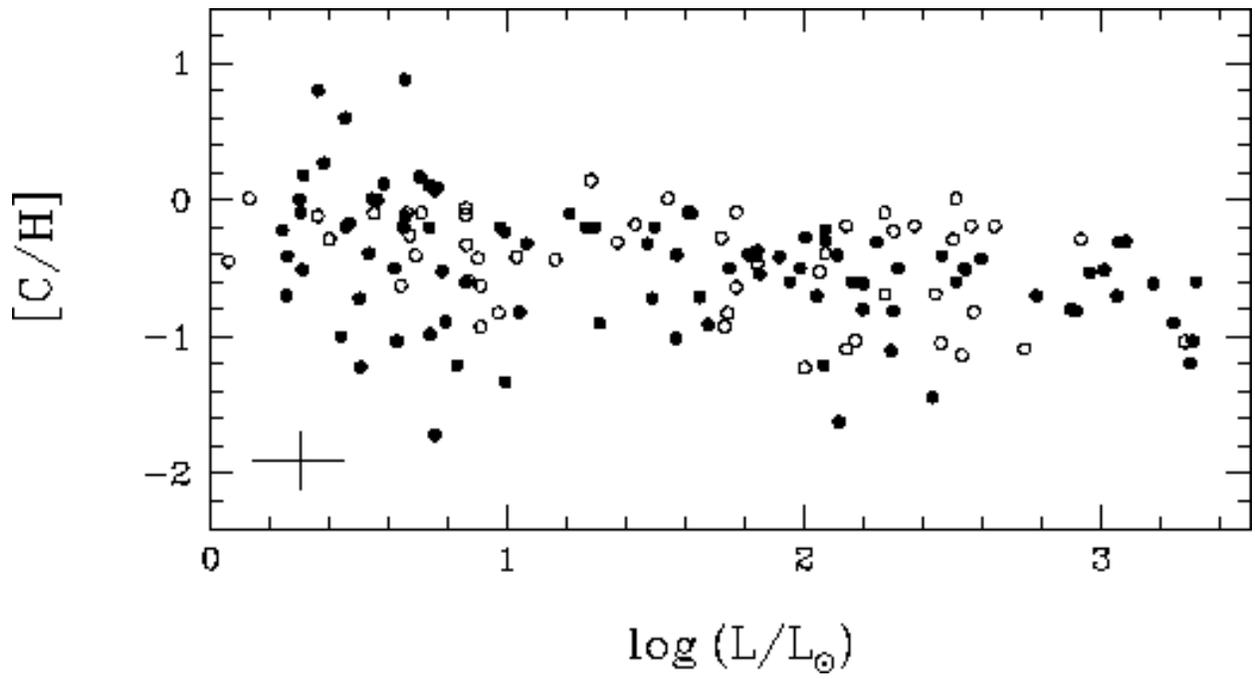}
\caption{The [C/H] as a function of luminosity for Ba-enhanced
         CEMP stars from \cite{aea07} (open circles) and for a larger sample of
         CEMP stars from \cite{lea06} (filled circles). The cross shows
         the data errorbars. Both samples have the same linear fit.
         }
\label{fig:f2}
\end{figure}

We begin with the normal halo star sample, and we use them to constrain our model of canonical extra mixing.
The sharp decline of [C/H] at $\log\,L/L_\odot\ga 2.2$ in the non-C-enhanced very metal-poor (VMP)  stars
(Fig.~\ref{fig:f1}) has been reproduced by us with
$r_{\rm mix} = 0.05\,R_\odot$ and $D_{\rm mix} = 0.1\,K$. The chosen depth is shown with the vertical black
line in Fig.~\ref{fig:f4}A. It guarantees that
material deficient in C is dredged up to the bottom of convective envelope, as indicated by observations.
A similar result can be obtained for a slightly smaller $r_{\rm mix}$
with an appropriately reduced value of $D_{\rm mix}$, e.g. for
$r_{\rm mix} = 0.045\,R_\odot$ and $D_{\rm mix} = 0.04\,K$. At the adjusted mixing depths
$T(r_{\rm mix})$ is lower than the temperature at the base of the H shell
by $\Delta\log T\approx 0.24$ and 0.19, 
respectively, which agrees very well with the estimates obtained by \cite{dv03} based on globular-cluster data.

\begin{figure}
\plotone{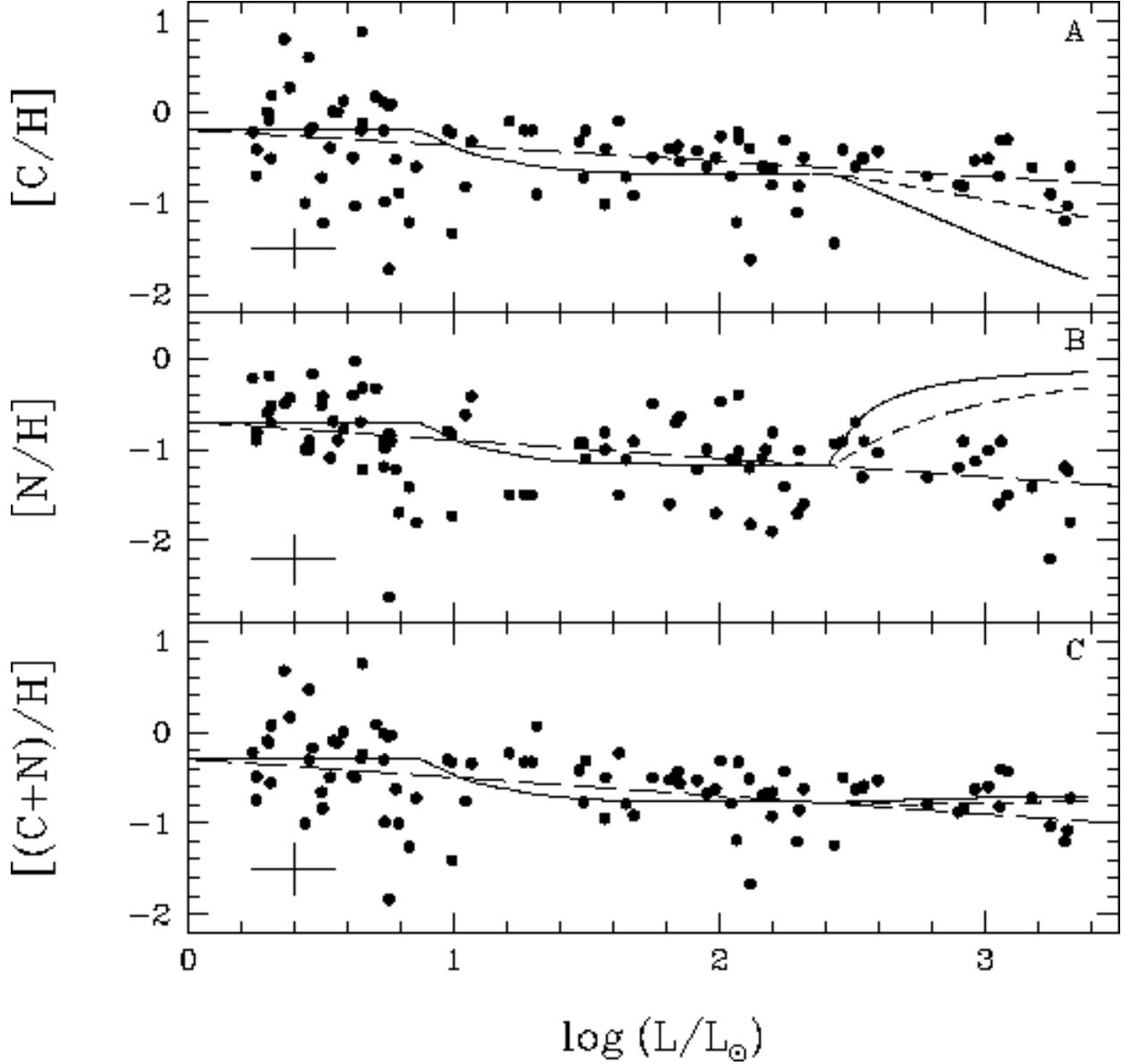}
\caption{The C and N abundances in the CEMP stars (circles) are compared with those predicted by
         our stellar evolution models in which canonical extra mixing has been parameterized with
         $r_{\rm mix} = 0.05$, $D_{\rm mix} = 0.1\,K$ (short-dashed red curves), and
         $r_{\rm mix} = 0.045$, $D_{\rm mix} = 0.04\,K$ (solid blue curves). Crosses show the observational errorbars.
         Black long-dashed lines are linear fits to the data with slopes
         $-0.18\pm 0.05$, $-0.21\pm 0.05$, and $-0.19\pm 0.04$ for [C/H], [N/H], and [(C+N)/H], respectively (\citealt{lea06}).
         } 
\label{fig:f3}
\end{figure}

\begin{figure}
\plotone{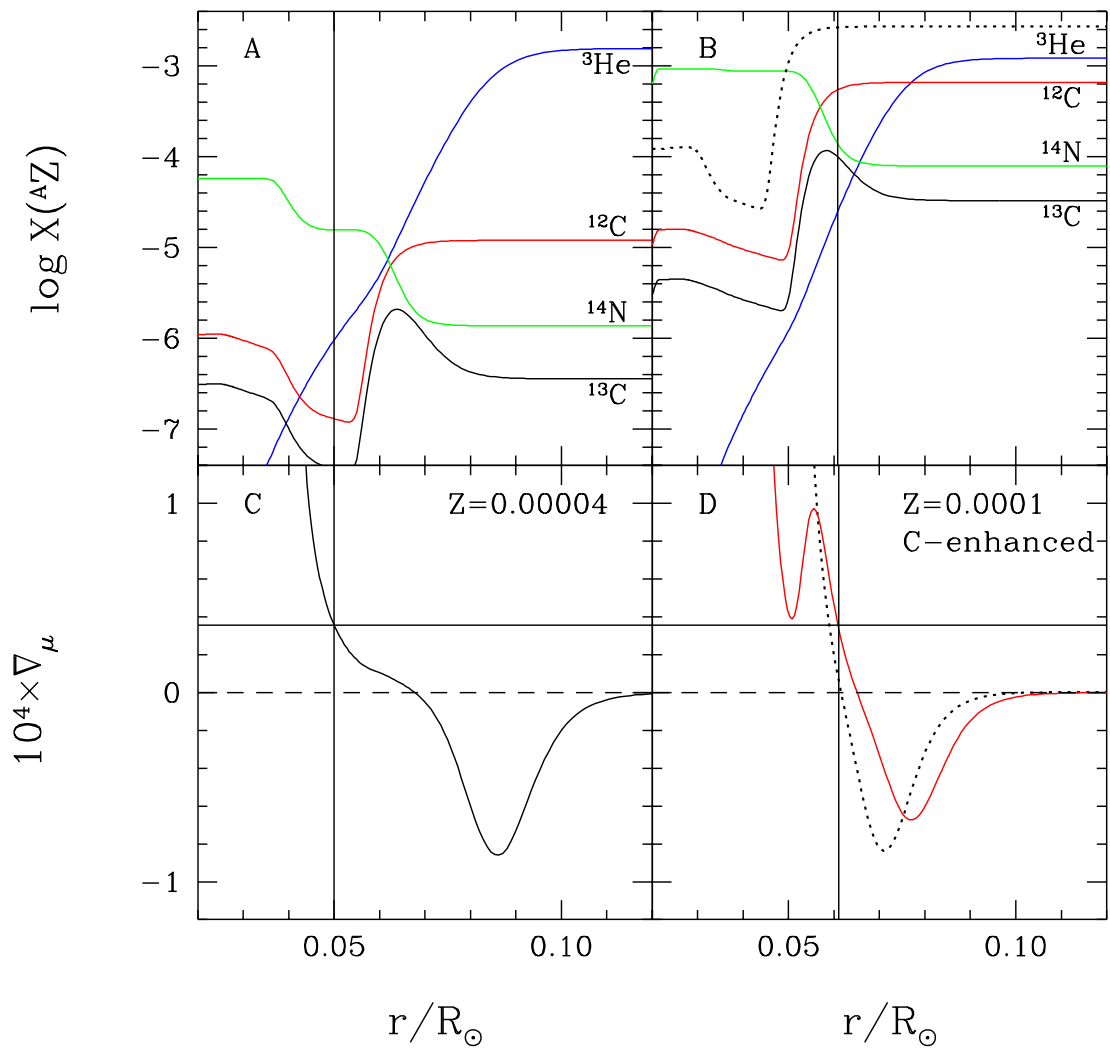}
\caption{The abundance (panels A and B) and $\mu$-gradient (panels C and D)
         profiles in our non-C-enhanced (panels A and C) and C-enhanced (panels B and D)
         upper RGB models. Dotted curves in panels B and D are from our solar-metallicity
         $1\,M_\odot$ upper RGB model.
         } 
\label{fig:f4}
\end{figure}

Our results for the CEMP stars are plotted in Fig.~\ref{fig:f3}.
The short-dashed red and solid blue curves have been computed for a stellar model with $Y=0.24$ and $Z=10^{-4}$
(the CEMP stars have $\langle[\mbox{Fe}/\mbox{H}]\rangle\approx -2.1$)
whose initial mass and CN abundances were $M = 0.7\,M_\odot$, [C/Fe]\,$= +0.3$, and [N/Fe]\,$= -0.5$. 
Following \cite{sea07}, we have assumed that
at the age of 0.86 Gyr the star began accreting material with the rate
$10^{-6}\,M_\odot$\,yr$^{-1}$ from its AGB binary companion. The total mass accreted and
its He and CN abundances were chosen to be $M_{\rm accr} = 0.15\,M_\odot$, $Y=0.34$, [C/Fe]$_{\rm accr} = +2.1$,
and [N/Fe]$_{\rm accr} = +1.6$.

Canonical extra mixing in our CEMP model has been modeled using the same
mixing parameters that we have adjusted for the non-C-enhanced VMP stars.
Comparing the [C/H] data at $\log\,L/L_\odot\ga 2.4$ with the short-dashed red and solid blue curves in Fig.~\ref{fig:f3}A, 
we infer that the efficiency of canonical extra mixing appears to be strongly reduced in the CEMP stars
compared to their non-C-enhanced VMP counterparts.

The reduced efficiency of canonical extra mixing in the CEMP stars becomes even more evident when we consider
their [N/H] values as a function of luminosity (Fig.~\ref{fig:f3}B).
CN processing for a high initial carbon should result in a sharp increase of [N/H] on the upper RGB (the short-dashed red and solid blue curves
at $\log\,L/L_\odot\ga 2.4$). Contrary to this, the observed trend of [N/H], as well as those of [C/H] and
the sum [(C+N)/H], with luminosity were characterized by \cite{lea06} as gradual declines (long-dashed black lines
in Fig.~\ref{fig:f3}). They have ascribed them to a dilution ``by mixing with a much larger
H content''. We concur in part, and interpret the decline from the MS to the RGB as being a dilution effect
produced by the FDU alone.  However, we also believe that representing this dilution as a continuing
process occurring on the RGB is not supported. The FDU is effectively complete at $\log\,L/L_\odot\approx 1.4$. 
At $1.4\la\log\,L/L_\odot\la 1.8$ the convective envelope mass is still increasing, but only by another 3\,--\,4\%.
So, no noticeable changes of the surface chemical composition due to the FDU dilution would be expected above 
$\log\,L/L_\odot\approx 1.4$, and we contend that no trend is seen.                             

Canonical extra mixing begins to operate only at $\log\,L/L_\odot\approx 2.4$ but it must {\em increase} the surface N abundance,
as it really does in non-C-enhanced VMP stars (the open blue circles in Fig.~\ref{fig:f5}), and not
{\em decrease} it. Hence, a dilution of the accreted He-, C-, and N-rich material can only occur
either on the MS, as a result of thermohaline convection, or on the lower RGB at $0.8\la\log\,L/L_\odot\la 1.4$,
as a result of the FDU. We will return to this problem in the next section. Here, we have to conclude
that, as revealed by the non-increase of [N/H] in the upper RGB CEMP stars, 
canonical extra mixing does not manifest itself in them. This conclusion agrees with the different
behavior of [C/H] at high luminosities in the CEMP and non-C-enhanced VMP stars (Fig.~\ref{fig:f1}).

Interestingly, there is another group of stars in which canonical extra mixing is observed to have a low efficiency.
These are low-mass RGB stars of nearly solar metallicity. Only modest depletions of
the carbon isotopic ratio ($^{12}$C/$^{13}\mbox{C}\approx 10$\,--\,15)
are reported in them, while the variations of [C/H] and [N/H] are close to         
their FDU limits (e.g., \citealt{gb91,oea06}). This contrasts with metal-poor red giants in which canonical extra mixing often reduces
the $^{12}$C/$^{13}$C ratio down to its equilibrium value $\sim$\,4, which is accompanied by the well-marked evolutionary decline
of [C/H]. \cite{chea98} have shown that such dependence of the efficiency of canonical extra mixing 
on metallicity is expected if the mixing cannot penetrate below some universal
critical $\mu$-gradient $(\nabla_\mu)_{\rm crit}$. Recently, \cite{dea06} have
proposed that the shallower mixing in the solar-metallicity red giants is caused by
a contribution to the $\mu$-gradient from the H depletion that accompanies
the C to N transformation. This contribution increases with the metallicity because $N(\mbox{C})\propto Z$.
At a metallicity close to the solar one, the H burning on the $^{12}$C nuclei builds up an additional hump in
the $\mu$-gradient profile that prevents mixing from approaching the H-burning shell as close as in metal-poor giants.
If this interpretation is correct then the reduced efficiency of canonical extra mixing in
RGB CEMP stars is almost definitely caused by their much higher C abundances compared
to their non-C-enhanced counterparts.      

To test this hypothesis, we confront the CN-abundance and $\nabla_\mu$ profiles in the vicinity of
the H-burning shell in our two upper RGB models in which the H shell has just erased the discontinuity
in the H-abundance profile left behind by the bottom of convective envelope at the end of the FDU
(it is not until this moment when canonical extra mixing begins to manifest itself). The first model represents the non-C-enhanced VMP stars:
it has $M=0.85\,M_\odot$, $Z=4\times 10^{-5}$, [C/Fe]\,=\,$+0.3$, and [N/Fe]\,=\,$-0.5$ (panels A and C in Fig.~\ref{fig:f4}).
The second had the initial mass $M=0.7\,M_\odot$ and abundances $Z=10^{-4}$, [C/Fe]\,=\,$+0.3$, and [N/Fe]\,=\,$-0.5$ but it
had accreted $0.15\,M_\odot$ of material with [C/Fe]\,=\,$+2.1$ and [N/Fe]\,=\,$+1.6$ (panels B and D in Fig.~\ref{fig:f4}).
The vertical line crossing panels A and C shows the mixing depth $r_{\rm mix} = 0.05$ that has been used to model  
the evolutionary decline of [C/H] in the non-C-enhanced VMP stars.
The horizontal line crossing panels C and D reads the value of $(\nabla_\mu)_{\rm crit}\approx 3.6\times 10^{-5}$ corresponding
to this mixing depth. When applied to the C-enhanced model this  $(\nabla_\mu)_{\rm crit}$
determines a shallower mixing depth ($r_{\rm mix}\approx 0.061$, panels B and D) with which extra mixing
will only be able to diminish the surface $^{12}$C/$^{13}$C ratio without changing [C/H] and [N/H]
(panel B). For comparison, dotted curves in panels B and D show the C-abundance and $\nabla_\mu$ profiles
in a solar-metallicity $1\,M_\odot$ RGB model. It is seen that in terms of the depth of canonical extra mixing the CEMP        
model is equivalent to the solar-metallicity model. Hence, it is most likely the mixing depth that is affected by
the variation of the star's C content in the first place. 

For our CEMP model we have intentionally chosen the large positive value of [N/Fe]$_{\rm accr}$ because 
it resembles the observed N abundance anomaly in unevolved CEMP stars.
It is thought that $^{12}$C dredged-up from the He-burning shell in AGB stars
might have partially been converted into N in the hot-bottom burning (in more massive
AGB stars) or by extra mixing similar to the canonical one (\citealt{nbw03,mea06,sea06,jea07}). 
To illustrate this point, we have reproduced in a slightly modified form Fig.~4 from the paper of \cite{sea06} in our Fig.~\ref{fig:f5}.
For non-C-enhanced VMP stars, we have used the data obtained by \cite{sea05} (the filled and open blue circles denote
{\it unmixed} and {\it mixed} RGB stars in their terminology). For CEMP stars, we have used the data compiled by
\cite{sea06} (the filled and open red squares represent stars with $\log\,g > 3.5$ and $\log\,g < 3.5$, respectively).
We have also plotted the C and N average abundances returned to the ISM by metal-poor AGB stars, as predicted by \cite{h04} (the magenta
asterisks with numbers below them giving the AGB model's mass).
The black curve in Fig.~\ref{fig:f5} shows changes of the surface C and N abundances
in an RGB model with $M=0.83\,M_\odot$ and $Z=0.0001$ produced by canonical extra mixing with the depth $r_{\rm mix} = 0.045\,R_\odot$ and
rate $D_{\rm mix} = 0.04\,K$. It is important to note that open blue circles ({\it mixed} stars) are located on
the upper RGB (\citealt{sea05}). This observational fact and a comparison of our black curve with the blue data points
support the conclusion made by \cite{sea05} that they had revealed the abundance patterns produced by canonical extra mixing in VMP stars.

It is interesting that most of the CEMP stars in Fig.~\ref{fig:f5} fill the gap between the C and N abundance yields predicted for the low-
and intermediate-mass AGB stars. The latter are well separated from the former because in the intermediate-mass AGB models
C is efficiently transformed into N at the bottoms of their convective envelopes (the hot-bottom burning). 
The mysterious location of the CEMP stars in the gap probably indicates that in their low-mass AGB donors
C was also transported down to a high-temperature region where it was partially transformed into N.
The rather low carbon isotopic ratios ($^{12}$C/$^{13}$C\,$< 40$) in low-luminosity s-element-rich CEMP stars
(\citealt{rea05}) support this hypothesis because without such extra mixing the low-mass AGB stars  
would produce $^{12}$C/$^{13}$C\,$> 1000$ (\citealt{h04}).

\section{Low Efficiency of Thermohaline Convection in MS CEMP Stars}

In the previous section, we have presented observational and theoretical arguments for 
the shallower depth of canonical extra mixing in RGB CEMP stars. The depth seems to be shallow enough
for extra mixing to leave the post-FDU C and N abundances unchanged. 
As we noted, the FDU starts at $\log\,L/L_\odot\approx 0.8$
and it effectively ends at $\log\,L/L_\odot\approx 1.4$
(the short-dashed red and solid blue curves in Fig.~\ref{fig:f3}). Hence, we
can compare the C and N abundance distributions for the CEMP stars with $\log\,L/L_\odot < 0.8$ and
for those of them with $\log\,L/L_\odot > 1.4$ to estimate the degree of dilution of the CN-rich material
by the FDU. These are plotted in panels A, B (for our larger sample), and C (for the smaller sample) in Fig.~\ref{fig:f6}. We see that the distributions for the stars
that have not experienced the FDU yet (shaded histograms)  are shifted toward higher abundances
with respect to the distributions for the stars in which the FDU has already ended (thick solid histograms). This holds true for
both the C (panels A, C) and N (panel B) abundance. 
There is a difference in the mean too, but it is not statistically significant because of the large star-to-star
scatter. However, there is a clearly defined median, and we employ a median based test.
We interpret the abundance distribution shifts as a signature of
the FDU dilution. The amplitude of the shift depends on both the average accreted mass $\langle M_{\rm accr}\rangle$ and the maximum mass of
convective envelope attained during the FDU $M_{\rm CE}^{\rm max}\approx 0.5\,M_\odot$.
To estimate their ratio, and hence $\langle M_{\rm accr}\rangle$, we shifted the shaded histograms to the left
by small steps in [X/H], where X stands for C and N (the shifts were applied to each star in the group), and we calculated
the Kolmogorov-Smirnov probability $P_{\rm KS}$ that the shifted shaded and thick solid histograms did not    
differ from one another after each step. Results of this statistical test are plotted in Fig.~\ref{fig:f6}D.
They show that the same shift $\Delta$[X/H]\,$\approx -0.4$ leads to the maximum and quite large values of $P_{\rm KS}$ for
both C and N. Neglecting the initial CN abundances, as compared to their accreted and post-FDU values,
we estimate $\langle M_{\rm accr}\rangle\approx 10^{\Delta [{\rm X}/{\rm H}]}\,M_{\rm CE}^{\rm max}\approx 0.2\,M_\odot$.

The fact that the pre-FDU and post-FDU C and N abundance distributions are surprisingly well separated implies
a low efficiency of thermohaline convection in MS CEMP stars, which has immediate implications for RGB mixing.
In Fig.~\ref{fig:f7}, we illustrate the $\mu$-profiles in our MS (red curve) and 
upper RGB (blue curve) CEMP models. On the scale of the MS inversion $(\Delta\mu/\mu)_{\rm MS}\sim 0.1$ the tiny RGB depression 
$(\Delta\mu/\mu)_{\rm RGB}\sim 10^{-4}$ at $r\approx 0.067\,R_\odot$ can hardly be seen. When comparing these profiles,
one should keep in mind that the efficiency of thermohaline convection is also proportional to the thermal diffusivity $K$ (equation \ref{eq:dmix}).
In the MS model, $K$ is three to four orders of magnitude lower than in the radiative zone of
the RGB model. As a result, the MS thermohaline diffusion coefficient can be less than 10\% of $D_{\rm mix}$ in the RGB model.
However, this rate is still fast enough to mix the MS CEMP star on a timescale shorter than 10 Myr, i.e. almost       
instantaneously compared to the star's MS lifetime. The evidence that this does not happen casts doubt on
thermohaline convection as the mechanism for canonical extra mixing. 

\section{Conclusion}

In this paper, we have put forward the hypothesis that the anomalously high C abundances in
the material accreted by MS CEMP stars
result in shallower depths of canonical extra mixing in these stars during their subsequent RGB evolution.
The shallower mixing depths are limited by steeper $\mu$-gradients produced by the H burning on abundant $^{12}$C nuclei.
A similar phenomenon occurs in open cluster RGB stars, which are of comparable total metal abundance.
Until now, it was not clear whether the reduced efficiency of mixing in open cluster giants was caused by their higher metallicity
or their higher mass relative to globular cluster giants.  We now see the same trend in low mass CEMP giants, and can therefore
conclude that metallicity is the primary determinant of the depth of mixing.  Mixing would still be expected in the outer envelopes of
these giants, and could impact other diagnostics, such as the $^{12}$C/$^{13}$C ratio.  Observational data on abundance trends
with luminosity of tracers of shallower mixing, such as the carbon isotopic ratio, would be a useful further test of this theoretical prediction.

For thermohaline mixing to be effective on the RGB the timescale must be 100 to 1000 times shorter than the \cite{kea80} estimates.
A comparable increase in efficiency on the MS relative to the \cite{sea07} estimates would imply a mixing timescale
far shorter than the MS lifetime.  As a result, we would
expect to see no dilution effect in the abundances, which contradicts the data.  We therefore conclude that thermohaline mixing is inefficient
in MS CEMP stars, contrary to the natural prediction of models where it is the dominant mechanism for RGB mixing.
In light of the potential importance of this conclusion, we now discuss the limits that we can place on
the timescale of MS mixing, potential observational and theoretical issues, and future observational tests.

We cannot make a similarly strong claim for thermohaline convection operating over Gyr timescales, as advocated by \cite{sea07}.
Our globally averaged result is a $0.2\,M_\odot$ average depth of the accreted material; this could be an unmixed $0.2\,M_\odot$ accreted layer,
or an initially lower mass accreted layer that has diffused to an average depth of $0.2\,M_\odot$.  A model with a diffusive timescale of order
the post-mass transfer MS lifetime and a $0.1\,M_\odot$ accreted layer, for example, would be consistent with our results.
When the mixing timescale is comparable to the evolutionary lifetime, a population synthesis approach is required for a comprehensive analysis.
One must account for the distribution of abundance anomalies in the donor stars, mass accretion amounts, and the range of secondary masses
over which the MS recipient can become a red giant.  Such calculations would be interesting, but are beyond the scope of this paper.

Observational selection effects are also a concern.  If there was a differential bias in the effective temperature estimates for dwarfs and giants,
for example, the relative abundance patterns could be impacted.  Similarly, a selection criterion where dwarfs would require higher carbon anomalies
than giants to be discovered would manifest itself as a dilution effect.  There is no evidence for the latter effect in the data; the observed carbon anomalies
are well-separated from the normal halo star pattern.  Effective temperature scale estimates are a more significant concern, especially for color-based
estimates, because of the spectral energy distribution distortions from very high [C/Fe] ratios.  \cite{aea07}, for example, found evidence for iron abundances
dependent upon the excitation potential of the lines, which is usually a signature of an incorrect effective temperature.  However, to impact our results there would
need to be a differential impact on the abundances derived in dwarfs and giants rather than an overall systematic effective temperature shift, and Aoki et al. did not
report such an effect.  We also note that very similar relative trends appear in samples where spectroscopic and color-based effective temperature estimates are
employed.

Is there a theoretical mechanism that could differentially suppress thermohaline convection in MS (but not RGB) stars?
\cite{dp07} have proposed that thermohaline convection can be suppressed by horizontal turbulent diffusion.
It is, at least in principle, possible to invoke stronger horizontal turbulence in dwarfs than in giants.
Angular momentum constraints from the existence of rapid rotation in horizontal branch stars do imply
strong differential rotation with depth in the convective envelopes of RGB stars
(\citealt{p83,pea92,sp00}).  One would therefore expect rotation rates
in the RGB envelope comparable to those in MS radiative interiors, and as a result comparable degrees of
horizontal turbulence.  This is an area that deserves further theoretical scrutiny.

\begin{figure}
\plotone{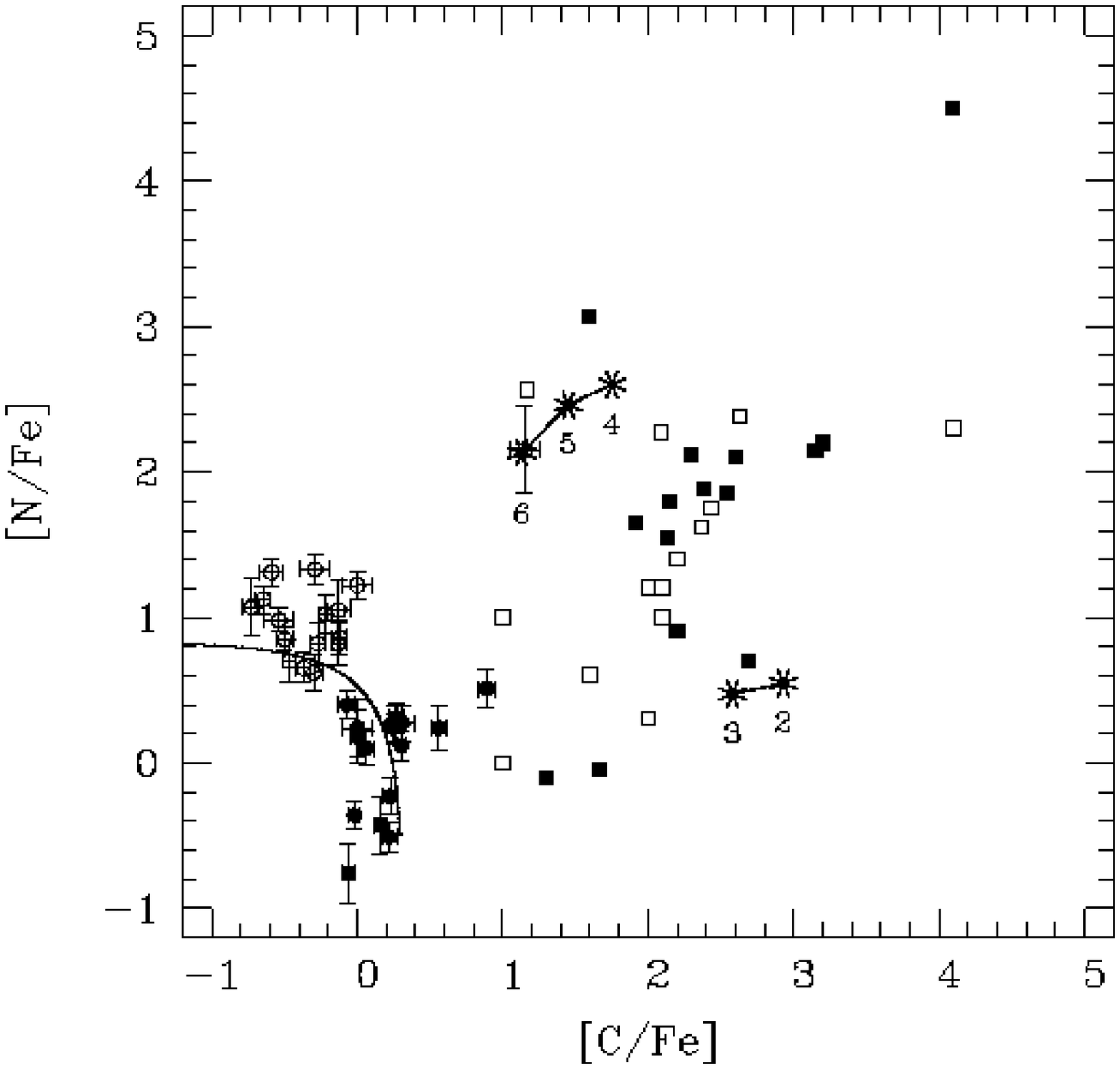}
\caption{The C and N abundances for the {\it unmixed} and {\it mixed} VMP stars (filled and open blue circles)
         obtained by \cite{sea05} as well as those for the CEMP stars with $\log\,g > 3.5$ and $\log\,g < 3.5$
         (filled and open red squares) compiled by \cite{sea06}. Magenta asterisks represent the average abundance yields
         predicted by \cite{h04} for metal-poor AGB models whose masses (in $M_\odot$) are given below the data points.
         Black curve shows changes of the surface C and N abundances produced by canonical extra mixing
         with the depth $r_{\rm mix} = 0.045\,R_\odot$ and rate $D_{\rm mix} = 0.04\,K$ in an RGB model
         with $M=0.83\,M_\odot$ and $Z=0.0001$. Note that most of the CEMP stars are located in the gap
         between the predicted AGB yields.
         } 
\label{fig:f5}
\end{figure}

\begin{figure}
\plotone{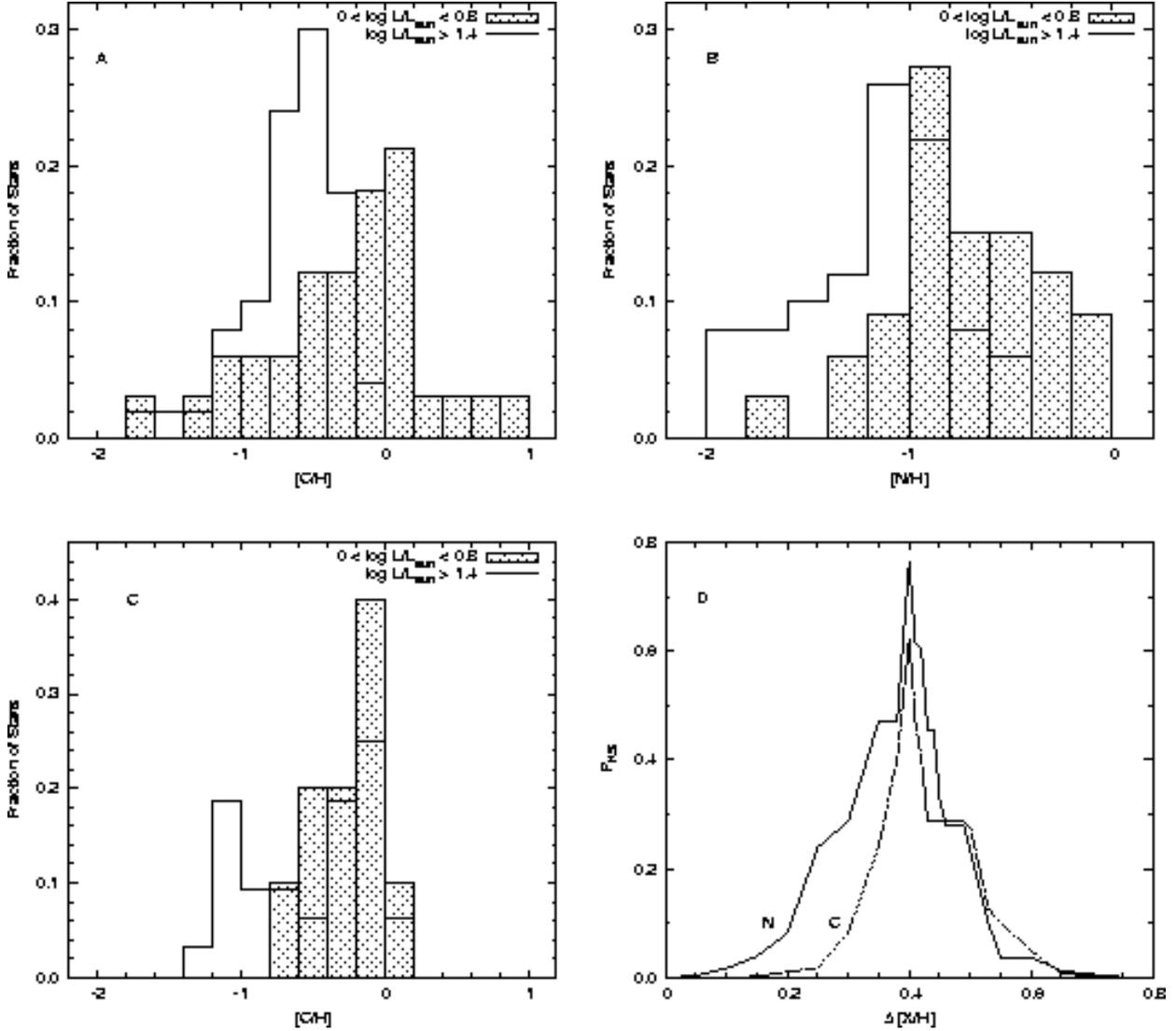}
\caption{The C (panels A and C) and N (panel B) abundance distributions for the CEMP stars
         that have not experienced the FDU yet (shaded histograms) and for those of them in which
         the FDU has already ended (thick solid curves). The data are from \cite{lea06} (panels A and B)
         and from \cite{aea07} (panel C). Panel D shows the Kolmogorov-Smirnov
         probabilities that the shaded histograms in panels A and B do not differ from their thick solid counterparts
         after being shifted to the left by $\Delta$\,[X/H].
         } 
\label{fig:f6}
\end{figure}

\begin{figure}
\plotone{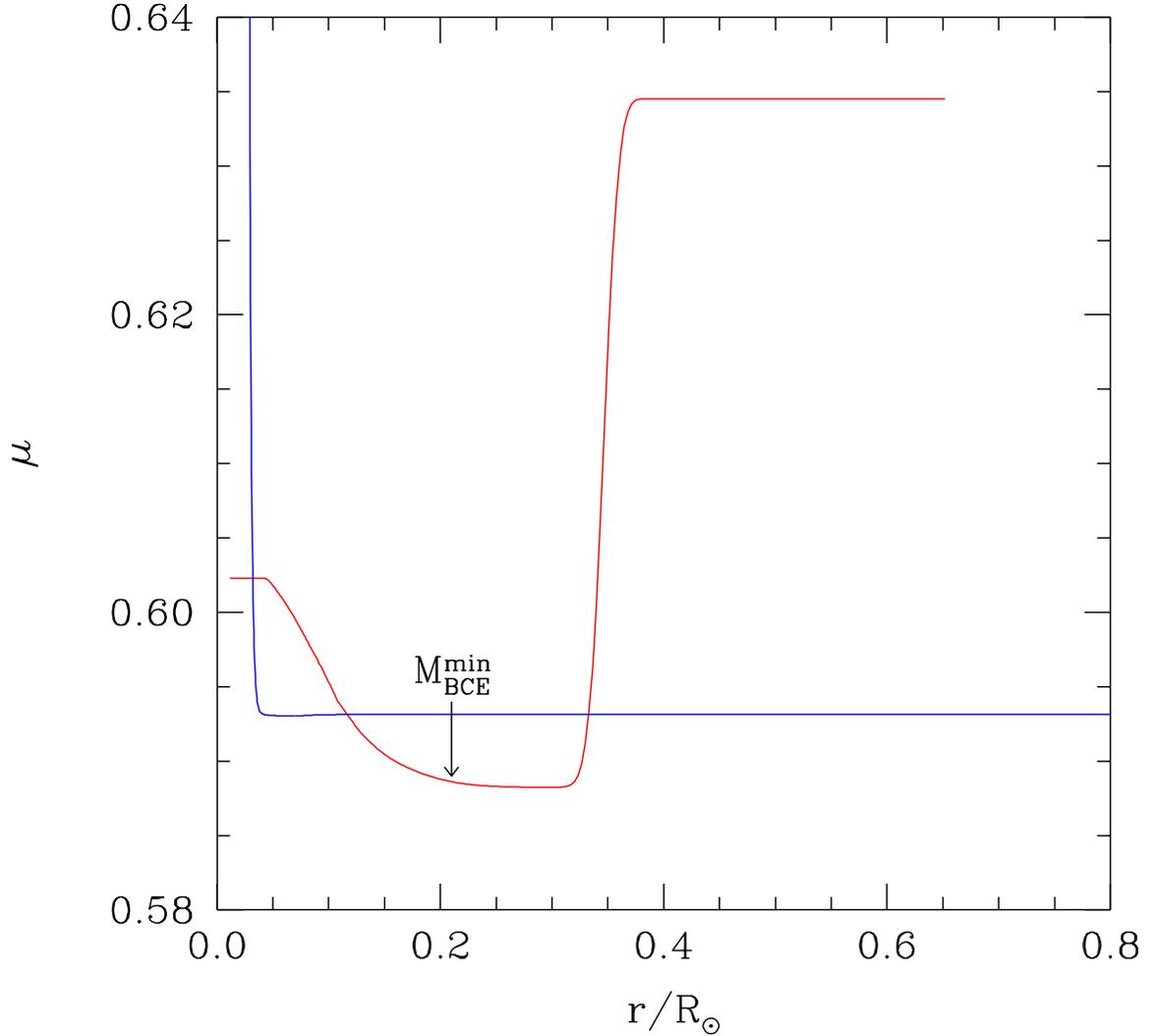}
\caption{The $\mu$ profiles in two models with $M=0.85\,M_\odot$ and $Z=0.0001$.
         The first (red curve) is our CEMP MS model that began its evolution
         as a $0.7\,M_\odot$ star and has just accreted $0.15\,M_\odot$ of AGB material
         enriched in He and C. The second (blue curve) is a model in which the H shell has just erased
         the H-abundance discontinuity. It has preserved its initial composition.
         The blue curve has a local minimum of
         $\Delta\mu/\mu\approx 1.4\times 10^{-4}$ located at $r\approx 0.067\,R_\odot$.
         It is produced by $^3$He burning.
         If canonical extra mixing in upper RGB stars is identified with thermohaline
         convection driven by this tiny $\mu$ inversion then a similar mechanism
         should produce very efficient mixing in CEMP MS stars as well, given their much
         stronger $\mu$ inversions (red curve), even though the thermal diffusivity is
         several orders of magnitude lower in the MS model (see text). 
         The arrow shows the radius of a mass shell
         down to which the base of convective envelope penetrates at the end of FDU.
         } 
\label{fig:f7}
\end{figure}

There is indirect evidence for extra mixing in low-mass AGB stars (e.g., see Fig.~\ref{fig:f5}
and its discussion in section 3). However, if extra mixing in RGB stars is really driven by $^3$He burning then it should
die out by the end of the RGB evolution because of the $^3$He exhaustion.
In this case, the $^3$He-driven thermohaline convection could not resume working
in the same stars on the asymptotic giant branch (AGB). So, we would expect
the absence of observational signatures of extra mixing in low-mass ($M\la 2\,M_\odot$)
AGB stars unless the mixing in them is of a different nature.
However, given the similarities in their depth and in the structure of radiative zone
where they operate, it is unlikely that the RGB and AGB mixing
have different physical mechanisms.

Further observational work would be extremely fruitful.  Samples should be examined for their binary
nature in order to generate binary mass transfer samples, and both the reliability of temperature estimates and
the relative and absolute abundance errors should be rigorously examined.  If thermohaline convection occurs over
an intermediate timescale, there should be a difference in the mean abundance level between MS stars that experienced
mass transfer recently and stars with older mass transfer episodes.  Hotter white dwarfs can be detected photometrically,
particularly for dwarf CEMP stars, and a difference in surface abundance between systems with younger and older
white dwarf companions would be definitive evidence for (or argument against) a moderate timescale for thermohaline mixing.

\acknowledgements
We are grateful to Sara Lucatello for sending us the C and N abundance data for the CEMP stars.
It is our pleasure to thank Jennifer Johnson, Thomas Masseron, Grant Newsham, Kristen Sellgren, and Don Terndrup for usefull discussions.
We acknowledge support from the NASA Grant No.\,NNG05 GG20G.


\end{document}